\documentclass[conference]{IEEEtran}

\usepackage{amssymb}
\usepackage{graphicx}
\usepackage[table,xcdraw]{xcolor}

\begin{document}


\title{Novel Architecture to Create and Maintain Personal Blockchains}

\author{\IEEEauthorblockN{Collin Connors}
\IEEEauthorblockA{\textit{Computer Science Department} \\
\textit{University of Miami}\\
Miami, FL \\
cdc104@miami.edu}
\and
\IEEEauthorblockN{Dilip Sarkar}
\IEEEauthorblockA{\textit{Computer Science Department} \\
\textit{University of Miami}\\
Miami, FL}
}

\maketitle

\begin{abstract}
Blockchain has been touted as a revolutionary technology. However, despite the excitement, blockchain has not been adopted in many fields. Many are hesitant to adopt blockchain technology due to privacy concerns, barriers to use, or lack of practical use cases. In this work, we outline a potential blockchain use case for tracking financial transactions across multiple financial institutions. We show the downsides of traditional centralized approaches and that blockchain approaches fail to give all the privacy and accessibility required for this use case. Thus we propose a novel blockchain architecture to support our use case. This novel architecture combines the ease of use of public blockchains with the privacy of private blockchains by allowing users to create personal blockchains. We believe this novel personal blockchain architecture will lead to more blockchain adoption, particularly in use cases handling private data.   
\end{abstract}

\section{Introduction}
\label{sec:Introduction}
People trust financial institutions to maintain records properly with little hesitation. These institutions are governed by various regulatory authorities and must comply with data security laws and regulations~\cite{BankingRegulationReview}. However, these financial systems are often independent of one another. Since a modern consumer or business uses many financial institutions, tracking all of their transactions can be challenging.

Individuals, for example, may have multiple credit cards issued by different banks, investment accounts, or other financial instruments. A business may have various incoming and outgoing invoices, utilize multiple banks, and even use various global financial institutions. Transactions across multiple financial institutions can be challenging to document. Thus false and fraudulent transactions are often missed even by the most vigilant consumers. 

An individual may mistakenly sign up for a subscription service and not notice charges for months. Alternatively, a business may pay fake invoices to malicious actors without realizing it. To spot these transactions, users must aggregate all of their transactions into a single ledger. Centralized applications such as Mint by Intui~\cite{intuitMintApplictation} help consumers generate this ledger. However, these applications can lead to unnecessary data privacy risks. Another proposed solution is to use blockchain technology similar to its use in cryptocurrency.

Cryptocurrency utilizes advanced technology for recording transactions of account holders~\cite{General_Blockchain_Description_2017}. Cryptocurrencies such as Bitcoin~\cite{NakamotoBitcoin} rely on digital ledgers known as a blockchain. A blockchain is a distributed, tamper-evident, tamper-resistant digital ledger~\cite{NISTBlockchain}. In its most basic form, a blockchain allows users to write data in an append-only manner without needing a centralized entity. 

As cryptocurrency has evolved, so has the way users interact with blockchains. While blockchain was initially designed to store cryptocurrency transaction records, modern blockchains allow storing varied and complex data. Developers can use platforms such as Ethereum~\cite{EthereumWhitepaper} to create dynamic distributed applications. Since security and accuracy are critical for financial applications, many developers utilize the blockchain to create decentralized FinTech applications~\cite{BDP_Features_Applications}. 

In this work, we propose a blockchain-based application that stores users' financial transaction history across multiple financial institutions on a blockchain. We show why centralized applications add unnecessary data privacy risks, and thus a blockchain-based solution is required. We then describe the issues if users store their financial transaction history on blockchains such as Ethereum. Thus, we propose a \textbf{novel} blockchain architecture for creating \textbf{personal blockchains} to overcome the issues posed by traditional blockchain applications. This work aims to highlight a need for the proposed novel architecture, and in future work, we plan to give a more detailed technical description of the architecture. 

We have organized the rest of this paper as follows. In section~\ref{sec:Background}, we provide a primer on blockchain technology and its uses. In section~\ref{sec:ProposedAppliction}, we provide a summary of our proposed application. In section~\ref{sec:NovelArchitectureToSupportTheApplication}, we expand on our summary providing details on how our novel blockchain architecture can be used to create the proposed application. Finally, we provide some concluding remarks and plans for future work in section~\ref{sec:Conclusion}.

\section{Background}
\label{sec:Background}
\par
Blockchains are  \emph{append-only}, \emph{tamper-evident}, \emph{tamper-resistant}, \emph{distributed fault-tolerant} digital ledgers~\cite{NISTBlockchain}. As the name suggests, a blockchain is a linked chain of blocks. A block consists of the block header and the block data. 

The block header contains the cryptographic hash of the previous block. Thus, any change to the previous block's header will indicate that it was tampered with. Likewise, the hash of the block data section is also included in the block header. Thus, if any change is made to the data, the hash value in the block header will not match the hash of the data. Therefore tampering with data in a block can be identified easily.

The block data contains a list of transactions. While initially, in Bitcoin, transactions were actual cryptocurrency transactions, modern blockchains allow a transaction to be any arbitrary data. For example, if a doctor's office was using a blockchain to track appointments, the transactions might be appointment dates and who made the appointment. 

A blockchain is a chronological ledger. The transactions are timestamped and ordered. Notice that transactions cannot be reordered after creating a block, as this would change the hash of the block data. Thus everything in the blockchain is chronological and cannot be changed. 

The first block of a blockchain is called the genesis block. This is the only block in the blockchain that's block header does not contain the cryptographic hash of the previous block. Because it has no previous block, it is critical to ensure that the creation of the genesis block is legitimate. For example, in a blockchain for cryptocurrency, one must ensure the creator of the cryptocurrency did not give themselves tokens in the genesis block. 

Next, let us briefly  review the concepts introduced in 1991 that laid foundation for \emph{append-only} \emph{tamper-evident}  electronic ledger~\cite{BitcoinCryptoIntoduciton}. To append a new document to an existing ledger, a digital digest of the previous document is added to the document. Then a digest of the document is digitally signed. Any change to the most recently appended document will change the digest of the document of the next documents, and tampering will be evident. Since a digital digest of the previous document has been included in the most recent document, any change to the previous document will not match with the digest on the most recent document, and tampering will be evident. This tamper-evident relationship between two consecutive documents is true for all documents in the ledger. 
 
Append-only tamper-evident and tamper-resistant electronic ledgers utilize cryptographic hash functions~\cite{cryptographicHashFnReview2012} and public-key cryptography~\cite{PublicKeyCryptography2009} to create digital digest and digitally signing the documents, respectively. Blockchains also utilize cryptographic hash functions and public-key cryptography for many purposes. For example, it is used for creating user identification numbers and signing transactions.  

Blockchains are decentralized. There is no central authority keeping track of the "real" blockchain; instead, many nodes worldwide agree on what transactions are in the blockchain. This method of agreeing is known as the Sybil Control mechanism~\cite{SybilControl}, sometimes referred to as the consensus protocol or consensus algorithm. There are many different Sybil Control mechanisms used by blockchains, such as Proof of Work, Bitcoin, and Proof of Stake, used by Ethereum. 

Blockchain networks are categorized based on their permission model~\cite{General_Blockchain_Description_2017}. Permissionless blockchains allow anyone to join the Sybil control mechanism and attempt to create new blocks. For example, Bitcoin is a permissionless blockchain network since anyone can try to mine blocks. In contrast, permissioned networks, such as those created by Hyperledger Fabric~\cite{HyperledgerFabricWhitepaper}, require permission from an authority to participate in the Sybil control.
Similarly, a blockchain ledger is categorized as public or private based on its viewing option. A \emph{public blockchain} ledger is available to everyone for viewing. Most permissionless blockchains are public ledgers because their contents are available to everyone for viewing. A \emph{private blockchain} ledger requires permission for viewing it.

\subsection{Limitation of Current Blockchain Ledgers}
\label{sec:LimitationOfCurrentBlockchainLedgers}
While blockchain technology can enhance many applications, the current blockchain architectures have limitations. Blockchain networks must choose between decentralization, scalability, and  privacy~\cite{Trilemma1997}. Thus no blockchain network is optimal in all three categories. Many blockchains chose to optimize decentralization at the cost of scalability and privacy. 

Likewise, permissionless blockchains have a cost when executing transactions. One such cost is the transaction fee users need to pay to use the network. While some blockchains have attempted to ensure low transaction fees, many users are still turned off. Likewise, users may need to wait for transactions to be approved. Some blockchains have attempted to speed up transaction times, but typically, these solutions are not scalable. 

Users who wish to avoid the drawbacks of permissionless blockchains may consider permissioned blockchains. However, these blockchains require technical expertise from the users to set up and maintain. Likewise, users need their own hardware to operate these blockchains, making them inaccessible to the average consumer.

We believe blockchain technology must be integrated with legacy technology, such as database technology. Moreover, a \emph{blockchain system} ought to be a collection of \textbf{\emph{independent services }}.  Different independent operators should provide these services. Thus, future users can deploy the blockchain system by choosing each service from a set of available service providers.

\section{Proposed Application}
\label{sec:ProposedAppliction}
\begin{figure}
	\centering
		\includegraphics[width=.45\textwidth]{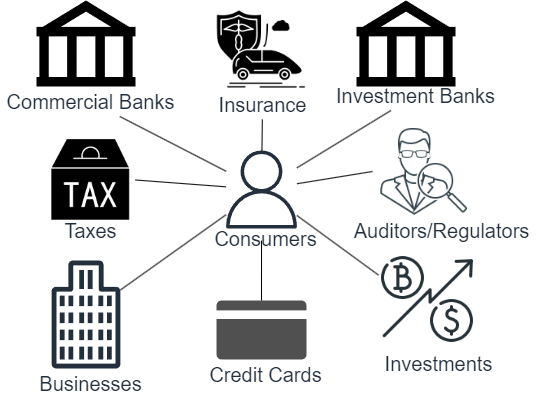}
	\caption{A high-level view of the traditional financial institutions. In this model, customers are responsible for collecting and maintaining all of their financial transactions.}
	\label{fig:Traditional}
\end{figure}

Financial systems have become more accessible and specialized. This specialization has led consumers to utilize multiple different financial intuitions. The complex web of various financial institutions has made it difficult for consumers to keep track of all their financial transactions. For example, an individual may have multiple banks, credit cards, and investments. Figure~\ref{fig:Traditional} shows how a customer interacts with existing financial institutions. The customer is receiving data from many different financial institutions. The consumer is responsible for collecting all of these sources of financial data, which can be time-consuming, tedious, and prone to mistakes. 

\begin{figure}
	\centering
		\includegraphics[width=.45\textwidth]{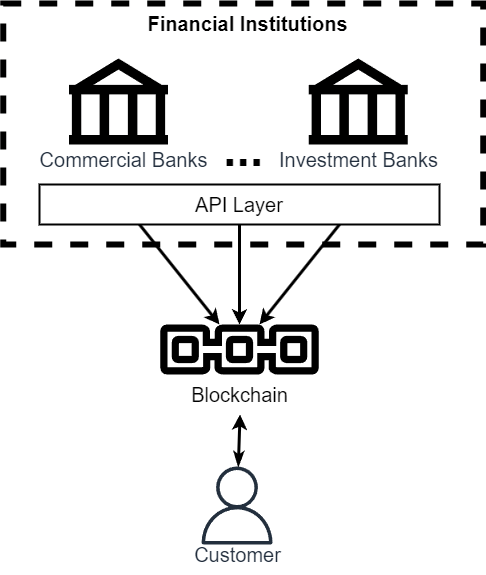}
	\caption{A high-level overview of the proposed blockchain-augmented financial application. This model stores a customer's financial transactions on a blockchain. We assume that the financial institutions already have an existing API layer to send transaction data.}
	\label{fig:HighLevel}
\end{figure}

To aid consumers, we propose an application that would aggregate all of their financial transactions across multiple financial institutions. Our proposed application relies on existing APIs provided by financial institutions to collect transaction data. Likewise, the application stores financial transactions in a blockchain. The blockchain gives users financial records immutability and chronological order, two highly desired properties when tracking financial records. Figure~\ref{fig:HighLevel} gives a high-level overview of such an application. In this model, all financial transactions are sent to the blockchain.  

Some centralized applications, such as Mint by Intuit~\cite{intuitMintApplictation}, have already attempted to solve this problem. These applications aggregate a user's financial transitions to the application. While this does make tracing complex financial transactions easier for consumers, these applications provide convince to the customer at the cost of privacy. Users must hand over their finical records to a centralized third party. This third party may use the financial transaction data for advertising or other purposes. Even if the third party does not utilize the data, if a hacker were to breach the third party, the customer's sensitive financial data could get leaked. Using a centralized application requires consumers to take on risk, which is undesirable in financial applications. 

Likewise, a similar application that allows users to store data on their own devices or cloud infrastructure does not provide all the desirable features. Critically past financial history should never change. Thus any storage should be immutable. Likewise, financial transaction history must be ordered. For example, consumers need to know if they have funds in their accounts before they make a purchase. Thus the consumer must ensure that they have funds deposited before they make a purchase. In addition, no party should have access to a consumer's financial records other than the consumer and the financial institutions. Considering these factors, blockchain presents itself as an ideal solution. However, current blockchain implementations are suboptimal for this proposed application.

If our proposed application were to use a permissionless blockchain such as Ethereum, users would have to publish their financial transactions to the blockchain. Since anyone can read a public blockchain, this could expose sensitive financial transactions. Even if the user were to take privacy measures such as encrypting the transactions, they would still risk leaking sensitive information. User error, mismanaged keys, or new technologies can reveal encrypted data. Lastly, users must pay a fee for every transaction in a permissionless blockchain. Even on blockchains with low fees, the cost can dissuade individuals and businesses alike from using our proposed application. 

In contrast, if our proposed application were to use an existing permissioned blockchain, such as Hyperledger Fabric, the user would be required to create and maintain their own blockchain infrastructure. Creating and maintaining infrastructure can be costly and complex. Likewise, users are responsible for managing permissions to the blockchain. If a user were to make a mistake with the permission set, they could grant access to parties who should not have access. A permissioned blockchain requires technical skill and computing power and is prone to mistakes making it a suboptimal solution for an application aimed at average consumers.

Thus we propose a \textbf{novel} blockchain architecture to create \textbf{personal blockchains} to support our proposed application. The following section describes the blockchain architecture and summarizes the modules necessary to support our application. This work aims to show the need for our novel blockchain architecture. We do not go into technical detail on our system but only show how such an architecture is ideal for our proposed application. 

\section{Novel Architecture to Support the Application}
\label{sec:NovelArchitectureToSupportTheApplication}
Since there is no existing blockchain development platform~\cite{BDP_Features_Applications} that supports our use case, we propose a \textbf{novel} blockchain architecture for creating \textbf{individual} blockchains. Our proposed architecture is unlike other blockchain architectures in that each user has their own blockchain rather than one single chain for all the users. Our proposed architecture aims to add more security and privacy to blockchain applications while being accessible to all users. 

Our novel architecture relies on modular, decentralized services to support the creation and maintenance of personal blockchains. In this work, we assume that these modules exist and are widespread. In future work, we will outline the technical specifications behind the modules and provide an implementation. For this work, it should be understood that each module is independent. Thus users could utilize multiple service providers and even change service providers during the application's lifecycle. 

While we describe our modules in the context of a financial application our architecture is general and can apply to many use cases. Similar to how web services such as hotmail first made email allowed the average person to create a personal email account, our proposed architecture can be used to bring personal blockchains to the average consumer. A single user may have multiple personal blockchains such as a blockchain for financial and a blockchain for health records. Our architecture is designed to make blockchain technology accessible to everyone regardless of technical knowledge.

\begin{figure}
	\centering
		\includegraphics[width=.45\textwidth]{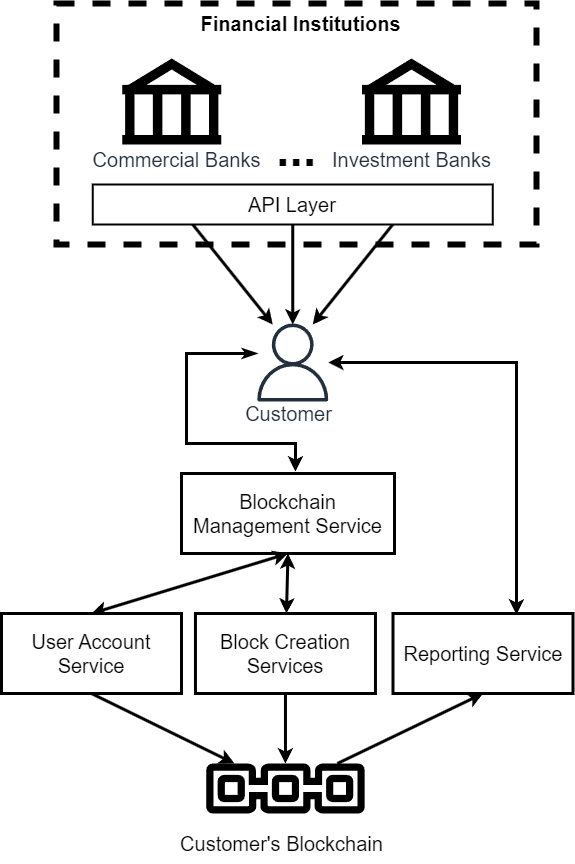}
	\caption{A block diagram of our novel architecture applied to our proposed application. In this model, the financial institutions send transactions, via an API layer, to the customer. The costumer then forwards those transactions to the various blockchain services. Note that each customer has their own individual blockchain.}
	\label{fig:BlockDiagram}
\end{figure}

Figure~\ref{fig:BlockDiagram} shows how our proposed architecture allows users to create personal financial blockchains. Through their APIs or other means, the financial institution will send data to the customer. Notice from the point of view of the financial institutions, nothing has changed, making this system easy to adopt. The customer forwards the data to the blockchain services. The blockchain services are made up of four independent modules:
\begin{enumerate}
	\item The Blockchain Management Service - Responsible for allowing access to the blockchain
	\item The User Account Service - Responsible for managing permissions on the blockchain
	\item The Block Creation Services - Responsible for adding new blocks to the blockchain
	\item The Reporting Service - Responsible for adding an application and presentation layer to the blockchain
\end{enumerate}
 We will briefly cover how the various components interact with our proposed application. We will give detailed technical specifications for each proposed module in future work. 

All our modules can be run decentralized, ensuring that no centralized entity can access users' data. The users can choose the extent of decentralization; some users may be comfortable using only centralized services, while others might prefer a fully decentralized approach. Likewise, users can choose how the blockchain is stored. Some may wish to store their blockchain on local storage or in the cloud. Others may prefer to use services such as IPFS to maintain decentralized storage. Lastly, users can decide the scope of transactions they wish to store. Some users may wish to store every transaction and all of its details. Other users may only store a summary of transactions. Since each user has their own personal blockchain, our architecture is highly flexible and can accommodate various users' concerns. 

\subsection{Blockchain Management Service}
\label{sec:BlockchainManagementSystem}

The Blockchain Management Service (BCMS) is responsible for allowing interactions between a user and their blockchain. This service provides an API for users to read and write data from a blockchain. The BCMS acts as an interface allowing our customers to read the blockchain's data and allow the various financial institutions to submit transactions to the blockchain. 

Likewise, the BCMS also provides access to blockchain storage. Customers can choose how they want their blockchain to get stored. Some customers may desire the blockchain to be stored on a device they control, such as a physical drive. Others may want to utilize centralized cloud storage options like Google Drive. Similarly, some users may wish to utilize decentralized storage such as IPFS. It is the job of the BCMS to ensure that regardless of the storage option, the user has access to their blockchain. 

Customers concerned about privacy could store the blockchain on a server they control, ensuring they are the only entity with a copy of their data. For example, a business cannot risk leaking sensitive financial transactions. Thus they could have the BCMS store the blockchain on an internal secured server allowing the business to maintain control of its data. 

This service is independent of the blockchain storage method. Thus users can change BCMS service providers without losing access to their blockchain. Additionally, since the BCMS does not store the blockchain itself, if a BCMS service provider were to be breached, users would leak far less data than in traditional centralized applications. 

\subsection{User Account Service}
\label{sec:UserAccountService}
The User Account Service (UAS) works with the BCMS to maintain customer and financial institutions' identities and permissions. The UAS acts like a certificate authority in web 2.0 or an identity management system in permissioned blockchains. The UAS issues certificates to customers and financial institutions containing their public keys, permissions, and other auxiliary information, such as the financial institution's API access tokens. 

Likewise, the UAS creates special blocks on the blockchain defining the permissions of users and financial institutions. For example, when customers start using our proposed application, they will need to tell the application which financial institutions they currently use and obtain API keys. The UAS is responsible for creating certificates for the financial institutions and writing to the blockchains to which institutions the customer has access. Storing the permissions on the blockchain allows other services to verify that the correct financial institutions are being queried. 

Similarly, throughout the life cycle of the proposed application, a customer may change financial institutions or join new financial institutions. It is the responsibility of the UAS to ensure that the proper permissions are maintained and reflect the institutions the customer currently utilizes. 

\subsection{Block Creation Services}
\label{sec:BlockCreationServices}
The Block Creation Services (BCS) are responsible for creating new data blocks and appending them to the user's blockchain. The BCS only has write access to the customer's blockchain, ensuring the customer's privacy. 

The BCS maintains a pending transaction queue for each customer. When the BCMS submits a new transaction to the BCS, it validates the transaction by authenticating the institution's and customer's digital signatures on the transaction. Then it inserts the current time stamp, digitally signs the transaction, and inserts it into the customer's pending transaction queue. 

Once a transaction queue accumulates sufficient transactions, the BCS creates a new block and publishes it by sending it to the \texttt{Blockchain Management Service} for storing the block. Two things are important to note: The number of transactions could be determined by the maximum size (say in MB) of a block or the end of a period, month, or quarter. Unlike mining nodes in a Bitcoin-like blockchain, the BCS does not store blockchains. The BCS only creates new blocks and is not responsible for storing and managing blockchain data.

\subsection{Reporting Service}
\label{sec:ReportingService}
The reporting service adds interconnection for the application and presentation layers to the user's blockchain. In contrast to the Block Creation Services, this service has read-only permissions. The read-only property prevents the reporting service from adding data when presenting data.

For all applications, the reporting service is responsible for checking the integrity of the blockchain. This service checks the hash of each block as well as the hashes stored in each block's header to ensure that the blockchain has not been tampered with. The reporting service reports to the user if any tampering has been detected.   

Likewise, this service is responsible for adding application features to our blockchain. In our proposed application, users will need to query their blockchain for specific data, such as all transactions made on a specific date. The reporting service runs these advanced queries and returns this information to the user. Similarly, this module allows for interaction between a frontend layer and the blockchain. The reporting service acts as an API that can query information from the blockchain and return it to a frontend. 

Notice that the reporting service's functionality is different for each application. For example, a healthcare application will require a different reporting service than a financial application. It is up to developers to create reporting service nodes for their desired applications. Likewise, the reporting service may be used to share blockchain data. For example, in a health care application a doctor may need to see a copy of health records stored on a users personal blockchain.

In this section, we only briefly describe the various components necessary for a novel architecture that supports our proposed application. This architecture is unlike other blockchain architecture as it allows for creating personal blockchains. The use of personal blockchains combines the ease of use and accessibility of permissionless blockchains with the privacy and control of data provided by permissioned blockchains. We plan to provide detailed technical specifications for our proposed novel architecture in future work. 

\section{Conclusion and Future Work}
\label{sec:Conclusion}
In this work, we described an application for tracking all of a customer's transactions across multiple financial institutions using blockchain. We discussed how traditional centralized approaches lead to a loss in data privacy. Likewise, we discussed why current blockchain development platforms would make such an application suboptimal. Thus we described a novel blockchain architecture to create personal blockchains. 

Our proposed architecture relies on four independent modules to create, maintain, and process the blockchain. Our proposed architecture blends the ease of use of public blockchains with the privacy-preserving features of private blockchains. Unlike traditional blockchains, where all users share one blockchain, our system allows each user to create their own personal blockchain. This novel architecture increases data privacy and security, allowing developers to create blockchain-based applications for fields that require high privacy. 

This work aimed to highlight a use case where traditional blockchain development platforms struggle to ensure privacy and accessibility. In future work, we plan to give an in-depth technical description of our proposed architecture. Likewise, we plan to implement our proposed architecture and our proposed application.

\bibliographystyle{plain} 
\bibliography{../References/blockchainWhitePapers,../References/cryptography,../References/surveys,../References/consensus,../References/generalBlockchain,../References/myPapers,../References/banking}
\end{document}